\documentclass[11pt]{article}
\begin{document}

\newfam\msbfam
\batchmode\font\twelvemsb=msbm10 scaled\magstep1 \errorstopmode
\ifx\twelvemsb\nullfont\def\Bbb{\bf}
        \font\fourteenbbb=cmb10 at 14pt
	\font\eightbbb=cmb10 at 8pt
	\message{Blackboard bold not available. Replacing with boldface.}
\else   \catcode`\@=11
        \font\tenmsb=msbm10 \font\sevenmsb=msbm7 \font\fivemsb=msbm5
        \textfont\msbfam=\twelvemsb
        \scriptfont\msbfam=\tenmsb \scriptscriptfont\msbfam=\sevenmsb
        \def\Bbb{\relax\expandafter\Bbb@}
        \def\Bbb@#1{{\Bbb@@{#1}}}
        \def\Bbb@@#1{\fam\msbfam\relax#1}
        \catcode`\@=\active
	\font\fourteenbbb=msbm10 at 14pt
	\font\eightbbb=msbm8
\fi
\catcode`\@=11
\def\Z {{\Bbb Z}}
\def\R {{\Bbb R}}
\def\E {{\Bbb E}}
\newfam\scrfam
\batchmode\font\twelvescr=rsfs10 at 12pt \errorstopmode
\ifx\twelvescr\nullfont
        \message{rsfs script font not available. Replacing with calligraphic.}
        \def\scr{\cal}
\else   \font\tenscr=rsfs10 
        \font\sevenscr=rsfs7
        \skewchar\twelvescr='177 \skewchar\tenscr='177 \skewchar\sevenscr='177
        \textfont\scrfam=\twelvescr \scriptfont\scrfam=\tenscr
        \scriptscriptfont\scrfam=\sevenscr
        \def\scr{\fam\scrfam}
        \def\cal{\scr}
\fi
\def\unit{\hbox to 3.3pt{\hskip1.3pt \vrule height 7pt width .4pt \hskip.7pt
\vrule height 7.85pt width .4pt \kern-2.4pt
\hrulefill \kern-3pt
\raise 4pt\hbox{\char'40}}}
\def\II{{\unit}}
\def\cM {{\cal{M}}}
\def\half{{\textstyle {1 \over 2}}}
\newcommand{\od}{\widetilde{\rm OD}}
\def    \beq    {\begin{equation}} \def \eeq    {\end{equation}}
\def    \bea    {\begin{eqnarray}} \def \eea    {\end{eqnarray}}
\def\la{\label} \newcommand{\eq}[1]{(ref{#1})}
\def    \lf     {\left (} \def  \rt     {\right )}
\def    \a      {\alpha} \def   \lm     {\lambda}
\def    \D      {\Delta} \def   \r      {\rho}
\def    \th     {\theta} \def   \rg     {\sqrt{g}} \def \Slash  {\, /
\! \! \! \!}  \def      \comma  {\; , \; \;} \def       \pl
{\partial} \def         \del    {\nabla}
\newcommand{\mx}[4]{\left#1\begin{array}{#2}#3\end{array}\right#4}
\newcommand{\Dpp}{\Delta + \nu}
\newcommand{\Dmm}{\Delta - \nu}
\newcommand{\Dp}{\Delta_+}
\newcommand{\Dm}{\Delta_-}
\newcommand{\Ds}{\left(\Delta^2-\nu^2\right)}
\newcommand{\Pp}{\Pi_+}
\newcommand{\Pm}{\Pi_-}
\newcommand{\lp}{\ell_p}
\newcommand{\ie}{{\em i.e., }}
\newcommand{\eg}{{\em e.g., }}
\newcommand\sss{\scriptscriptstyle}
\newcommand\scs{\scriptstyle}
\newcommand{\bc}{\begin{center}}
\newcommand{\ec}{\end{center}}
\newcommand{\nz}{\normalsize}
\newcommand\nn{\nonumber}
\frenchspacing
\begin{titlepage}
\begin{flushleft}
       \hfill                      {\tt hep-th/yymmxxx}\\
       \hfill                        G\"oteborg-ITP-preprint\\
       \hfill                                  UG-01-32\\
\end{flushleft}
\vspace*{1.8mm}
\begin{center}
{\bf\LARGE Deformation independent open brane metrics and generalized 
theta parameters}\\
\vspace*{4mm}
{\large David S. Berman\footnote{berman@racah.phys.huji.ac.il}}\\
\vspace*{2mm}
{\small Racah Institute for Physics \\ 
 Hebrew University,  Jerusalem 91904, Israel}\\
\vspace*{3mm}
{\large
Martin Cederwall, Ulf Gran, Henric Larsson,  \\
Mikkel Nielsen, 
Bengt E.W. Nilsson\footnote{ martin.cederwall, gran, 
solo, mikkel, tfebn@fy.chalmers.se}}\\
\vspace*{2mm}
{\small Institute of Theoretical  Physics\\
G\"{o}teborg University and Chalmers University of Technology\\
SE-412 96 G\"{o}teborg, Sweden}\\
\vspace*{3mm}
{\large and Per Sundell\footnote{p.sundell@phys.rug.nl, 
address from 1 September 2001: Department of Theoretical Physics, Uppsala 
University.}}\\
\vspace*{2mm}
{\small Institute for Physics, University of Groningen \\
\small Nijenborgh 4, 9747 AG Groningen, The Netherlands}\\
\vspace*{3mm}
\end{center}
\noindent \underbar{\bf Abstract:}
\noindent We investigate the consequences of generalizing certain well 
established properties of the open string metric to the 
conjectured open membrane and open D$p$-brane 
metrics. By imposing deformation independence on these metrics their 
functional dependence on the background fields can be determined including
the notorious conformal factor. In analogy with the non-commutativity 
parameter $\Theta^{\mu\nu}$ in the string case, we also obtain 
`generalized' theta parameters which are rank $(q+1)$ antisymmetric tensors 
(polyvectors) for open D$q$-branes and rank 3 for the open membrane case. The 
expressions we obtain for the open membrane quantities are expected to be 
valid for
general background field configurations, while the open D-brane quantities 
are only  
valid for one parameter deformations. By reducing the open membrane data to five 
dimensions, we show that they, modulo a subtlety
with implications for the relation between OM-theory and NCYM, 
correctly generate the open string and open D2-data.

\end{titlepage}

\section{Introduction}

Open string metrics play an important role in the study of open
strings in non-trivial backgrounds. In particular, non-commutative
Yang-Mills (NCYM)  \cite{con}-\cite{Delle} and non-commutative open
string theories (NCOS) \cite{gopa}-\cite{solo1} can arise on the world
volume of D-branes, probing backgrounds with non-zero NS-NS two-form
potentials. For constant background fields on the world volume of the
D-brane a straightforward derivation of the two-point function between
two string coordinates reveals that the closed string metric and NS-NS
two-form potential get replaced by an effective open string metric and a 
$\Theta$-parameter governing the
non-commutativity of world volume coordinates. We also get an effective open string
 coupling. These open string quantities determine in these backgrounds  
all the physical properties of the D-brane such as the open string spectrum. 

When studying the physics of D-branes, one is often interested in
describing the brane decoupled from the bulk closed string theory. This
essentially means taking a scaling limit of the closed string
quantities. One may then use the expressions of the open string
quantities given in terms of the closed string data to determine the
fate of the open strings on the D-brane \cite{GMSS}. 

Similar decoupling limits have been studied not only in the 
case of open strings ending on D-branes, but
 also for the case of open membranes ending on
five-branes in eleven dimensions
 \cite{GMSS}-\cite{davidper} and
various situations with open D-branes 
 ending on other D-branes
\cite{lu,soloper} or 
 NS5-branes \cite{GMSS,harmark2}-\cite{obers1}.

In these other cases, involving the open membrane or other open $p \geq 2$ 
branes, we do not know how to proceed with quantization.
However, instead of trying to extract important quantities, for, \eg open 
membranes, from 
$n$-point functions one may try to determine some general principles or key 
properties, 
which will make it possible to uniquely  construct the 
open membrane analog of the open string metric and 
non-commutativity (or theta) parameter as done in \cite{us2}. The tensor structure
of the metric (\ie the metric modulo its conformal factor) 
could be determined from the equations of motion on
the five-brane, while the overall conformal factor could be fixed, 
to leading order, by
demanding that in the decoupled theory the dimensionless open membrane
metric be fixed \cite{us2}. This was checked (in the limit) using IIB/M-theory
duality. 
Further evidence for the scaling behavior of the conformal factor
was put forth in \cite{davidper}, where it was shown that it implies that
the near 'horizon' region of the self-dual string soliton on the 
five-brane decouples from the 'bulk' five-brane theory in a Maldacena-like
scaling limit of the five-brane world volume theory, and that this 
region has an interesting fixed OM-metric geometry.

The complete conformal factor was recently 
derived by requiring that the open membrane metric 
should reduce to the open string metric when reducing 
to ten dimensions \cite{janpieter}. The derivation in that work is based on
an assumption regarding the relation between the open string coupling
and the component of the open string metric in the compactified direction, 
similar to how the string coupling arises from the eleven dimensional metric.
One of the primary goals of this paper is to use an alternative method to
determine the overall conformal factors of the various open brane
metrics, based on non-trivial properties that one extrapolates from
knowledge of the open string metric. It is then possible to check
consistency, as we will show below in this paper, by examining the
 dimensional reduction of such quantities
and see if they match to known results.

The central idea that will be used here is the property of deformation
 independence (described in detail below)
 \cite{intr,peet,russo23,Berman} to obtain explicit expressions for
 the open brane metrics and generalized theta ($\Theta$) parameters for the
 cases
 mentioned above.  These theta parameters generalize the non-commutativity
parameter $\Theta^{\mu\nu}$ appearing in the open string case
to higher dimensional open branes. However, since here the connection to 
non-commutativity is
unclear we will only refer to them as theta parameters. Theta parameters
of rank larger than two have also been discussed in \eg\ \cite{others}. 
One might, however, 
speculate that these open brane 
quantities also for  $q \geq 2$  contain some
information about the two-point and $(q+1)$-point functions of
the deformed quantum theory.

Starting with the observation that deformation
 independence is a property of the open string metric, we postulate that this
 principle extends to other brane metrics. 
Although we do not explore this avenue
here, it may in fact be possible to derive these open brane metrics
using various kinds of dualities. As will be clear in the later sections
some of our arguments  supporting the implementation of 
deformation independence for open branes 
are closely related to such duality arguments. For instance, T-duality relates
as usual open D$q$-brane quantities to open D($q\pm1$)-brane quantities.

The paper is organized as follows.
After explaining the concept of deformation independence in
section two, we derive in section three the
 conformal factor for our proposed open D-brane metrics. In this section we also
 construct generalized theta parameters for open D$q$-branes,
based on the idea that they must be scale independent. In section four
we then repeat this exercise for the open
 membrane. By demanding deformation independence, we find 
 an open membrane metric that
 agrees with the one obtained in \cite{janpieter}
 by reducing the open membrane metric to
 open string theory and by demanding that it correctly produces OM theory
in the decoupling limit defined in  \cite{us2}. Section four also contains
a derivation, showing that the open membrane theta parameter is linearly 
self-dual with respect to the open membrane metric.
In section five we check that the eleven dimensional quantities, 
the open membrane metric and theta parameter, 
correctly reduce to the expressions for the open string and open D2-brane 
in ten dimensions. Some concluding 
remarks are given in section six.

\section{Open string data and deformation independence}

The data governing the effective open string perturbation theory
on a D$p$-brane in a closed string background with string frame
metric $g_{MN}=g_{\mu\nu}\oplus g_{ij}$, dilaton $e^{\phi}$ and
two-form potential $B_{\mu\nu}$ are given by the open string
two-point function \cite{sw}\footnote{Our definition of the
non-commutativity parameter $\Theta$ differs from the one in ref.
\cite{sw} by a factor $2\pi$.} and the effective open
string coupling. In this paper we use ten-dimensional spacetime
indices $M=0,\dots,9$, $(p+1)$-dimensional world volume indices
$\mu=0,\dots,p$, and $(9-p)$-dimensional transverse space indices
$i=p+1,\dots,9$. The two-point function is

\begin{eqnarray}  <T[X^{\mu}(\tau) X^{\nu}(0)]>&=&  -\alpha'
G^{\mu\nu}\log{|\tau|} + i\pi\Theta^{\mu\nu}\epsilon(\tau)\
,\nonumber
\\  <T[X^i(\tau) X^j(0)]> &=& -\alpha' g^{ij}\log{|\tau|}\ ,
\label{ncos2pt}
\end{eqnarray}
where
\beq \label{osd1}
G_{\mu\nu}=g_{\mu\nu}+B_{\rho\mu}g^{\rho\sigma}B_{\sigma\nu}\ ,
\eeq \beq \label{osd2}
\Theta^{\mu\nu}=-\alpha'g^{\mu\rho}B_{\rho\sigma}G^{\sigma\nu}\ .
\eeq
The symmetric tensor $G_{\mu\nu}$ is the open string metric governing the
 mass-shell
condition for the open string states propagating on the D-brane, and
the antisymmetric tensor $\Theta^{\mu\nu}$ is
the parameter of non-commutativity between the D-brane coordinates. 
The open string coupling constant is given by \cite{sw}

\begin{equation}
G^{2}_{{\sss \rm{OS}}}= e^{\phi}\left(\frac{\det G}{\det
g}\right)^{1/4}\ .
\end{equation}

It is important to note that, as will be explained below, the open string 
metric and coupling are invariant under
deformations of the tensor fields in the supergravity background
\cite{Berman,Delle} that correspond to non-commutative
deformations of a D-brane worldvolume (open string) theory. Such
a supergravity solution is generated by a stack of $N$ D-branes
and corresponds to a U(N) gauge theory. Higgsing it to
$U(N-1)\times U(1)$ by giving a VEV to one of its scalars,
corresponds to separating off one of the branes from the stack.
Given that we are in a large $N$ limit, the supergravity solution is
unaffected and the separated brane that contains the U(1) degrees
of freedom acts like a probe on the dual spacetime. The statement
of deformation independence is then that the coupling and
effective metric of these U(1) degrees of freedom as a function
of the Higgs value are independent of non-commutative deformations
of the field theory.

The non-commutativity parameter implies a deformation of the
underlying algebra of functions, in terms of which the worldvolume
theory is defined. From this point of view it is clear that it
should not depend on the energy scale of the probe. This can be
verified explicitly using the supergravity dual description of the
probe brane, \ie the open string non-commutativity parameter given
above does not depend on the location of the probe brane in the
dual spacetime \cite{Berman,Delle}.

The deformation independence implies that for both electric and
magnetic deformations we have \cite{Berman}

\begin{eqnarray}\label{osmc}
G_{\mu\nu}|_{\rm deformed}&=&G_{\mu\nu}|_{\rm undeformed}~
=g_{\mu\nu}|_{\rm undeformed}~=H^{-\frac{1}{2}}\eta_{\mu\nu}\ ,\nonumber\\
G_{{{\sss \rm OS}}}^{2}|_{\rm deformed}&=&G_{{{\sss \rm
OS}}}^{2}|_{\rm undeformed}~= e^{\phi}|_{\rm
undeformed}~=gH^{\frac{3-p}{4}}\ ,\end{eqnarray}
where $H$ is a harmonic function.
The NCOS (electric NS-NS two-form deformation) and NCYM (magnetic
NS-NS two-form deformation) supergravity duals are obtained by taking
the following near horizon limits ($\alpha^\prime\rightarrow 0$):

\begin{equation} {\rm NCOS}\quad :\qquad
\frac{x^{M}}{\sqrt{\alpha'}}\ ,\quad g\quad{\rm fixed}\ ,
\end{equation}
\begin{equation}{\rm NCYM}\quad :\qquad
x^{\mu}\ ,\quad \frac{x^{i}}{\alpha'}\ ,\quad g^{2}_{\rm YM}=
g(\alpha')^{\frac{p-3}{2}}\quad \quad{\rm fixed}\ .
\end{equation}
The NCOS limit amounts to keeping fixed the ten
scalars of the two-dimensional worldsheet theory, while the NCYM
limit keeps fixed the transverse scalars of the probe brane gauge
theory. Thus the former `sigma-model' limit leads to the supergravity dual of a
non-commutative open string theory, while the latter
`field-theory' limit yields the dual of a non-commutative Yang-Mills theory,
even though (\ref{osmc}) is valid in both cases (see \eg
\cite{Berman}).

Supergravity solutions corresponding to D$p$-brane bound states
can be obtained in many different ways, leading to equivalent
solutions related by coordinate changes. In particular, by acting with 
transformations in the T-duality group $O(p+1,p+1)$ \cite{Delle,Berman}, which
act on the supergravity fields while keeping the spacetime coordinates fixed,
the deformation independence of the open string metric and coupling constant becomes 
manifest, as written in (\ref{osmc}).
Other deformation methods that involve deformation dependent coordinate
transformations do not lead to manifest deformation independence.
Hence there exists a special choice of coordinates which
implies (\ref{osmc}). 
In these preferred coordinates the near horizon limits are of
sigma-model and field theory type in the case of electric and
magnetic NS-NS two-form deformations, respectively. The converse is not true, 
however, since there are deformation dependent reparametrizations 
that do not affect the nature 
of the near horizon limit while they
upset (\ref{osmc}), \ie the preferred coordinates cannot 
be identified uniquely by examining the
near horizon limits alone. In order to use the requirement of deformation 
independence to determine the open string metric, one has to identify the 
above-mentioned $O(p+1,p+1)$ action.
 
In \cite{soloper} analogous preferred coordinates appropriate to open
D$q$-branes on D($q+2$)-branes were identified by acting first with
 S-duality and then with T-duality transformations on the (D3,F1) bound state. 
In \cite{soloper} it was shown that  
these coordinates indeed yield near horizon limits which are of sigma model
 type (\ie all ten coordinates $x^{M}$ scale homogeneously in the near horizon
 limit) in the case of electric deformations of the RR $(q+1)$-form
potentials, and of field theory type in the case of magnetic such
deformations. 

Assuming that the open D-brane metrics are
deformation independent (in these coordinates), we can use
the form of the supergravity solutions to determine the covariant
form of these quantities as functions of the closed D-brane data.
Similarly, the generalized
non-commutativity (or theta) parameters can be determined by requiring 
independence of energy scale. For simplicity, we shall apply these ideas in
 the next section in the case of minimal rank deformations.

The above approach can be `lifted' to M-theory. The appropriate
coordinates are given in section four 
below. As a result 
we find the deformation independent open membrane metric and scale independent 
theta-parameter as covariant expressions of the background 
three-form tensor field.
These expressions are general since the three-form deformation in M-theory 
is general, even though the reductions to string theory may be restricted.

\section{Open brane metrics and generalized theta parameters}

\subsection{The open D$q$-brane metrics}

In this subsection we will obtain the open D$q$-brane metric that is 
valid for a background deformed by only one RR field, a $(q+1)$-form.  
As discussed above, we commence with  the 
assumption that deformation independence is also valid for open D$q$-branes,
 similar to the open string case. 
We will obtain the open D$q$-brane metric in
the case of an electrically deformed D$(q+2)$-brane background, corresponding
 to the $\widetilde{\rm{OD}q}$-theories, which are obtained in the decoupling
 limit of open D$q$-branes ending on D($q+2$)-branes \cite{soloper}, and then
 show that it also holds in the case of a magnetically RR deformed background,
 corresponding to the generalized non-commutative gauge theories D$q$-GT of 
\cite{soloper}, as well as for NS5-brane
 backgrounds, corresponding to the OD$p$-theories. 

The relevant part of the supergravity solution, corresponding to a
D$(q+2)$-brane with an electric RR ($q+1$)-form turned on,
is given by \cite{soloper}\footnote{Note that here $\theta$ is dimensionless,
 while in \cite{soloper} $\frac{\theta}{\alpha'}$ was dimensionless.}

\bea \label{ODq}
ds^{2}_{q+3}&=&(Hh_{-})^{-{1\over2}}
dx_{q+1}^2+(H^{-1}h_{-})^{\frac{1}{2}}dx_{2}^{2}\ ,\nonumber\\
e^{2\phi}&=&g^{2}h_{-}^{3-q\over 2}H^{{1-q\over2}}\ ,\nonumber\\
C_{01\cdots q}&=&-{\theta\over gHh_{-}}\,,\quad B_{q+1,q+2}=-\theta H^{-1}\\
h_{-}&=&1-\theta^{2}H^{-1}\,,\quad H=1+\frac{gN(\alpha')^{\frac{5-q}{2}}}
{r^{5-q}}\ ,\nonumber \eea 
where $dx_{q+1}^2=-dx^{2}_{0}+\cdots +dx_q^2$ and
$dx^2_{2}=dx^2_{q+1}+dx^2_{q+2}$. 
To obtain this solution we have 
started with an electrically NS-NS deformed D3-brane 
(in the particular coordinates discussed in the previous section),
 \ie a (D3,F1) bound 
state. Then we have used S-duality followed by T-duality to obtain the 
(D($q+2$),D$q$) bound state. 
It is now natural to assume that since we choose to start from a
 (D3,F1) bound state, given in coordinates such that we get a 
deformation independent open string metric, then after S-duality the new 
(D3,D1) bound state should give a 
deformation independent open D1-string metric \cite{mikkel}.
Finally, applying T-duality should provide us with solutions,
giving  deformation independent open D$q$-brane metrics. 

The open D$q$-brane metric will now be derived by considering open D$q$-branes 
ending on D($q+2$)-branes. The expressions will also be correct for open 
D$q$-branes ($q\neq 2$) ending on NS5-branes, as will be shown below. 
We will make the following Ansatz for the open D$q$-brane metric, 
which is similar in structure to the open string metric (\ref{osd1}):
\begin{equation}\label{ansatz}
G^{{\sss \widetilde{\rm{OD}q}}}_{\mu\nu}=A(X)g^{{\sss \rm{D}q}}_{\mu\nu}+
\frac{1}{q!}B(X)(C^{2}_{q+1})_{\mu\nu}\ ,
\end{equation}
where $A(X)$ and $B(X)$ are functions of 
\begin{equation}
X=\frac{1}{(q+1)!}C_{q+1}^{2}\ ,
\end{equation}
$g^{{\sss {\rm D}q}}_{\mu\nu}=e^{-{2\phi\over q+1}}g_{\mu\nu}$ is the
closed D$q$-brane metric, and 

\beq \label{C2}
(C^2)_{\mu\nu}=g^{\rho_1\sigma_1}_{{\sss {\rm D}q}}\cdots g_{{\sss {\rm D}q}}
^{\rho_q\sigma_q}C_{\rho_1\dots\rho_q\mu}C_{\sigma_1\dots\sigma_q\nu}\
,\quad C_{q+1}^2=g_{{\sss {\rm D}q}}^{\mu\nu}(C^2_{q+1})_{\mu\nu}\
.\eeq 
This Ansatz contains only one potential $C_{\mu_1...\mu_{q+1}}$, 
corresponding to the fact that we only consider one-parameter deformations.
$A(X)$ and $B(X)$ can now be obtained by inserting the background solution 
(\ref{ODq}) and demanding deformation independence. This implies that 
the open D$q$-brane metric $G^{{\sss \widetilde{{\rm OD}q}}}_{\mu\nu}$ 
 is equal to the closed D$q$-brane metric
in the original undeformed configuration, 
\beq g^{{\sss {\rm D}q}}_{\mu\nu}|_{\rm 
undeformed}=(g^2H)^{-{1\over q+1}}\eta_{\mu\nu}\ .\la{cldqm1}\eeq
This leads to two equations for $A$ and $B$, one in the $2$
directions perpendicular to the electric flux, and one in the
parallel directions (note the sign due to the lorentzian
signature):

\begin{equation} A(X)(g^{2}H)^{-\frac{1}{q+1}}h_{-}^{\frac{q-1}{q+1}}=
(g^2H)^{-{1\over q+1}}\ ,
\end{equation}
\begin{equation}
A(X)(g^{2}H)^{-\frac{1}{q+1}}h_{-}^{-\frac{2}{q+1}}-B(X)(g^{2}H)^{-\frac{1}
{q+1}}h_{-}^{-\frac{2}{q+1}}\theta^{2}H^{-1}=(g^2H)^{-{1\over q+1}}\
.\end{equation}
To obtain these equations we have used that 
\begin{eqnarray}
g^{{\sss {\rm D}q}}_{\alpha\beta}&=&(g^{2}H)^{-\frac{1}{q+1}}h_{-}^{-\frac{2}
{q+1}}\eta_{\alpha\beta}\ ,\nonumber\\
g^{{\sss {\rm D}q}}_{ab}&=&(g^{2}H)^{-\frac{1}{q+1}}h_{-}^{\frac{q-1}{q+1}}
\delta_{ab}\ ,\nonumber\\
\frac{1}{q!}(C^{2}_{q+1})_{\alpha\beta}&=&-\theta^{2}H^{-\frac{q+2}{q+1}}
(gh_{-})^{-\frac{2}{q+1}}\eta_{\alpha\beta}\ , \\
\frac{1}{q!}(C^{2}_{q+1})_{ab}&=&0\ ,\nonumber\\
X&=&-\theta^{2}H^{-1}\ ,\nonumber
\end{eqnarray}
where $\alpha,\beta=0,\ldots,q$ and $a,b=q+1,q+2$.

Solving for $A(X)$ and $B(X)$ from the above algebraic
equations we find
\beq A(X)=B(X)=(1+X)^{1-q\over 1+q}\ ,\la{ab}\eeq  
which gives the following open D$q$-brane metric
\beq\label{ODqmetric} G^{{\sss \widetilde{ \rm{OD}q}}}_{\mu\nu}=
 \Big[1+{1\over
(q+1)!}C_{q+1}^2\Big]^{{1-q\over
 1+q}}\left(g^{{\sss {\rm D}q}}_{\mu\nu}+{1\over
q!}(C^2_{q+1})_{\mu\nu}\right)\ .\eeq 

Note that for $q\neq 1$ these open brane 
metrics are only expected to be valid for a one parameter
 deformation, \ie only one RR field is turned on.
For $q=1$ we find
agreement with the expected result for the open D-string \cite{mikkel}, which we expect to be valid for any 
deformation, while for $q=2,3$ we have the
following open D$2$-brane and D$3$-brane metrics:

\bea G^{{\sss \widetilde{{\rm OD}2}}}_{\mu\nu}&=& 
\Big[1+{1\over 6}C_3^2\Big]^{-{1\over
3}}\left(g^{{\sss {\rm D}2}}_{\mu\nu}+{1\over
2}(C_{3}^{2})_{\mu\nu}\right)\ ,\nonumber \\
G^{{\sss \widetilde{{\rm OD}3}}}_{\mu\nu}&=&
\Big[1+{1\over 24}C_4^2\Big]^{-{1\over
2}}\left(g^{{\sss {\rm D}3}}_{\mu\nu}+{1\over
6}(C_{4}^{2})_{\mu\nu}\right)\ .\label{OD23}\eea

As a first 
check of the expressions for the metrics we evaluate (\ref{ODqmetric}) in a 
magnetically (instead of electrically) RR deformed background \cite{soloper}.
 The relevant part of this solution is
\bea\label{dm}
ds^2_{q+3}&=&h_+^{\frac{1}{2}}H^{-\frac{1}{2}}(-dx_0^2+dx_1^2)+h_+^{-\frac{1}
{2}}H^{-\frac{1}{2}}(dx_2^2+\cdots+dx_{q+2}^2)\,,\\
e^{2\phi}&=&g^2H^{\frac{1-q}{2}}h_+^{\frac{3-q}{2}}\,,\qquad 
C_{2\ldots q+2}=\theta g^{-1}H^{-1}h_+^{-1}\,,\nn
\eea
where $h_+=1+\theta^2 H^{-1}$. 
This gives (\ref{cldqm1}), which is the expected (deformation independent)
 result.

We can also check (\ref{ODqmetric}) in the case of a NS5-brane, deformed by an
 electric RR $(q+1)$-form, $q\neq 2$, where the relevant part of the solution 
is \cite{soloper}
\bea\label{NS5e}
ds^2_{1+5}&=&h_-^{-\frac{1}{2}}(-dx_0^2+\cdots+dx_p^2)+h_-^{\frac{1}{2}}
(dx_{p+1}^2+\cdots+dx_{5}^2)\,,\\
e^{2\phi}&=&\tilde{g}^2H h_-^{\frac{3-p}{2}}\,,\qquad 
C_{01\ldots p}=-\theta \tilde{g}^{-1}H^{-1}h_-^{-1}\,,\nn
\eea
or a magnetic RR $(q+1)$-form, $q\neq 2$, where the relevant part of the 
solution is \cite{soloper}
\bea\label{NS5m}
ds^2_{1+5}&=&h_+^{\frac{1}{2}}(-dx_0^2+\cdots+dx_{4-q}^2)+h_-^{-\frac{1}{2}}
(dx_{5-q}^2+\cdots+dx_{5}^2)\,,\\
e^{2\phi}&=&g^2H h_+^{\frac{3-q}{2}}\,,\qquad 
C_{5-q\ldots 5}=-\theta g^{-1}H^{-1}h_+^{-1}\,.\nn
\eea
Also in these cases we get
 (\ref{cldqm1}), which is the expected deformation independent result.

In the case $q=2$ both $C_{012}$ and $C_{345}$ 
are non-zero due to the duality relation. This means that (\ref{ODqmetric}) is 
no longer valid (see also the final paragraph of section 5) and has to be replaced
 with the open membrane 
metric (\ref{OMmetric}) below, where we of course must use the closed D2-brane
 metric instead of the closed M2-brane metric. This can easily be realized
 from the fact that the closed D2 and M2-brane metrics are 
`identical'\footnote{Note that the harmonic functions are not the same, since
 the NS5-brane has four transverse directions while the M5-brane has five.}, 
and 
that $C_{\mu\nu\rho}=\tilde{g}^{-1}A_{\mu\nu\rho}$. Note also that 
$C_{\mu\nu\rho}$ obeys 
the same self-duality relation as $A_{\mu\nu\rho}$. This self-duality has to 
be satisfied in order to obtain the correct number of degrees of freedom.
 It is also because of this self-duality constraint that we obtain a result 
which is different from the na\"\i ve open D2-brane metric given by 
(\ref{OD23}).

Next, we will take an electric near horizon limit of the 
solution (\ref{ODq}), which gives the `relevant' part of the supergravity 
dual of $\widetilde{{\rm OD}q}$. We do this in order to obtain the 
$\widetilde{{\rm OD}q}$ length scale $\ell_{{\sss \widetilde{{\rm OD}q}}}$, 
which we will then compare with the expression given in \cite{soloper}. 
The electric near horizon limit is defined by keeping \cite{soloper}
\begin{equation}\label{ell}
\tilde{x}^{\mu}=
\frac{\tilde{\ell}}
{\sqrt{\alpha'}}x^{\mu}\ , \tilde{r}=\frac{\tilde{\ell}}
{\sqrt{\alpha'}}r\ , \theta=1\ , g\ , \tilde{\ell}\ ,
\end{equation}
fixed in the $\alpha'\rightarrow 0$ limit.
This limit gives the following open D$q$-brane metric
\begin{equation}
\frac{G^{{\sss \widetilde{{\rm OD}q}}}_{\mu\nu}}{\alpha'}=\frac{1}
{\ell_{{\sss \widetilde{{\rm OD}q}}}^{2}}H^{-\frac{1}{q+1}}\ ,
\end{equation}
where $\ell_{{\sss \widetilde{{\rm OD}q}}}=g^{\frac{1}{q+1}}\tilde{\ell}$. This
 calculation gives exactly the same expression for the  
$\widetilde{{\rm OD}q}$ length scale as in \cite{soloper}, which we expected.
Note also that the metric is finite in the UV limit 
($\tilde{r}\rightarrow\infty$, $H\rightarrow 1$ ). Finally, the tension
of the open D$q$-branes is fixed in the decoupling limit and can be calculated
 from\footnote{Note that we ignore factors of $2\pi$.}
\beq
T= \Big(\frac{\sqrt{-{\rm det}g^{{\sss {\rm D}q}}}}{\alpha'^{\frac{q+1}{2}}}
-\epsilon^{01\ldots q}\frac{C_{01\ldots q}}{\alpha'^{\frac{q+1}{2}}}\Big)\,.
\eeq
Inserting the solution (\ref{ODq}), we get the following tension in the 
UV limit 
\begin{equation}
T=\frac{1}{2\ell_{{\sss \widetilde{{\rm OD}q}}}^{q+1}}\ .
\end{equation}

\subsection{Generalized theta parameters}

Next we will obtain an expression for a generalized theta
 parameter for a theory of open D$q$-branes. Again these expressions will only
 be  valid for a one parameter deformation (except for $q=1$). In 
the string case, 
$\Theta^{\mu\nu}$ is given by the antisymmetric part of the two-point function,
 which should be expressed in terms of the background fields. When we have a 
$C_{q+1}$-form deformation with $q\neq 1$, we cannot construct an antisymmetric
 two-index tensor bilinear in the background fields, but one can instead try
 to write down quantities with more than two antisymmetrized indices that might
in fact arise as an
antisymmetric part of a $(q+1)$-point function.  
 Following the open 
string case where\footnote{Note that this definition differ with a factor of 
$2\pi$ from the usual definition \cite{sw}.}
\begin{equation}\label{OStheta}
\Theta^{\mu\nu}_{{\sss {\rm OS}}}=-\alpha'g^{\mu\rho}B_{\rho\sigma}
G^{\sigma\nu}\ ,
\end{equation}
we make the following Ansatz for the generalized theta
 parameter
\begin{equation}\label{ansatz2}
\Theta^{\mu_{1}\cdots \mu_{q+1}}_{\widetilde{{\sss {\rm OD}q}}}=
-(\alpha')^{\frac{q+1}{2}}
\tilde{A}(X)g^{\mu_{1}\nu_{1}}_{{\sss {\rm D}q}}
C_{\nu_{1}\ldots \nu_{q+1}}G^{\nu_{2}\mu_{2}}_{{\sss \widetilde{{\rm OD}q}}}\cdots
G^{\nu_{q+1}\mu_{q+1}}_{{\sss \widetilde{{\rm OD}q}}}\ ,
\end{equation}
where $\tilde{A}(X)$ is a function of $X=\frac{1}{(q+1)!}C_{q+1}^{2}$. Now using 
(\ref{ODq}) and the assumption that $\Theta^{0\ldots q}_{\widetilde{{\sss 
{\rm OD}q}}}=-\ell_{{\sss \widetilde{{\rm OD}q}}}^{q+1}$ (we choose the minus 
sign in order 
to have a minus sign in (\ref{ODqtheta}) below, similar to (\ref{OStheta})),
 when we have taken the electric near horizon limit (\ref{ell})
\footnote{Note also 
that this assumption implies that $\Theta^{0\ldots q}_{{\sss {\rm OD}q}}=
-\theta g(\alpha')^{\frac{q+1}{2}}$, before one has taken the electric near 
horizon limit.},
 gives
\beq \tilde{A}(X)=(1+X)^{-\frac{1-q}{1+q}}\ .\la{a'}\eeq 
The assumption above is motivated by the fact that the theta parameter in
 the NCOS case is constant and equal to the open string length scale 
\cite{intr,peet,Berman}, \ie $\Theta^{01}_{{\sss {\rm OS}}}=\alpha'_{\rm eff}$.
This yields the following generalized theta parameter 

\begin{equation}\label{ODqtheta2}
\Theta^{\mu_{1}\cdots \mu_{q+1}}_{\widetilde{{\sss {\rm OD}q}}}=
-(\alpha')^{\frac{q+1}{2}}
(1+\frac{1}{(q+1)!}|C_{q+1}|^{2})^{-\frac{1-q}{1+q}}g^{\mu_{1}\nu_{1}}
_{{\rm D}q}C_{\nu_{1}\ldots \nu_{q+1}}G^{\nu_{2}\mu_{2}}_{{\sss \widetilde{{\rm OD}q}}}
\cdots G^{\nu_{q+1}\mu_{q+1}}_{{\sss \widetilde{{\rm OD}q}}}\ .
\end{equation}

In the case of a one parameter solution, this theta parameter can also be 
written as
\begin{equation}\label{ODqtheta}
\Theta^{\mu_{1}\cdots \mu_{q+1}}_{\widetilde{{\sss {\rm OD}q}}}=
-(\alpha')^{\frac{q+1}{2}}
(1+\frac{1}{(q+1)!}C_{q+1}^{2})^{\frac{1-q}{1+q}}g^{\mu_{1}\nu_{1}}
_{{\sss {\rm D}q}}\cdots g^{\mu_{q}\nu_{q}}_{{\sss {\rm D}q}}
C_{\nu_{1}\ldots \nu_{q+1}}G^{\nu_{q+1}\mu_{q+1}}_{{\sss \widetilde{{\rm OD}q}}}\ .
\end{equation}
In fact, we have $q+2$ possible expressions with the number of 
$g^{\mu_i\nu_i}$-s ranging from zero to $q+1$. The reason is that in the case
 of  a one parameter solution, it can be 
written in terms of an SO(1,$q+2$)/SO($1,q)\times$SO(2) parametrization for an electric 
$C_{q+1}$ (similarly we have an SO(1,$q+2$)/SO(1,1)$\times$SO($q+1$) parametrization for 
a magnetic $C_{q+1}$ and in this case the formulas below hold when $\alpha, 
\beta$ and $a, b$ are exchanged)
\beq\label{c2p}
(C^2_{q+1})_{\alpha\beta}=\frac{1}{q+1}\,C^2_{q+1}\,g^{{\sss 
{\rm D}q}}_{\alpha\beta}\,,\qquad (C^2_{q+1})_{ab}=0\ .
\eeq
Inserting this in the open D$q$-brane metric yields
\beq
G^{{\sss \widetilde{{\rm OD}q}}}_{\alpha\beta}=
(1+\frac{1}{(q+1)!}C_{q+1}^{2})^{\frac{2}
{q+1}}g^{{\sss {\rm D}q}}_{\alpha\beta}\,,\qquad G^{{\sss \widetilde{{\rm OD}q}}}_{ab}=
(1+\frac{1}{(q+1)!}C_{q+1}^{2})^{\frac{1-q}{q+1}}g^{{\sss {\rm D}q}}_{ab}\ ,
\eeq
and we can therefore replace a $g_{{\sss {\rm D}q}}^{\alpha\beta}$ with a 
$G_{{\sss \widetilde{{\rm OD}q}}}^{\alpha\beta}$, but then we lose a factor of
 $(1+X)^{\frac{2}{q+1}}$. 

It is not obvious that the expressions (\ref{ODqtheta2}) and 
(\ref{ODqtheta}) are totally antisymmetric, but we can show 
that this indeed is the case.  In the M2-brane case we have a three-form 
which obeys a non-linear self-duality relation on the M5-brane. 
Using dimensional reduction and T-duality, we get a relation between 
the ($q+1$)-form and the two-form on the D($q+2$)-brane
\beq
G_{{\sss \widetilde{{\rm OD}q}}}^{\mu_1\nu} C_{\nu}{}^{\mu_2\cdots\mu_{q+1}}\sim
\epsilon^{\mu_1\cdots\mu_{q+3}}B_{\mu_{q+2}\mu_{q+3}}\ .
\eeq
We thus see that (\ref{ODqtheta}) is proportional to the lefthand 
side of the relation above, and this theta is therefore totally 
antisymmetric. In general, the expressions for the theta parameters 
above will contain higher odd powers of $C_{q+1}^{\alpha_1\cdots\alpha_{q+1}}$. 
Since (\ref{ODqtheta}) only contains the powers one and three, 
we see that $(C_{q+1}^3)^{\alpha_1\cdots\alpha_{q+1}}$ is totally 
antisymmetric. In order to show the antisymmetry for (\ref{ODqtheta2}), 
we only need to show that three is the highest power even in this case. 
To see this we use the following relation (which can be obtained by 
dimensional reduction of the M5-brane) on the D4-brane for the three-form 
\beq
(C_3^4)_{\mu\nu}=\frac{1}{3}C_3^2\, (C_3^2)_{\mu\nu}\ .
\eeq
T-dualizing this expression yields
\beq
(C_{q+1}^4)_{\mu\nu}=\frac{1}{q+1}\,C_{q+1}^2\,(C_{q+1}^2)_{\mu\nu}\ ,
\eeq
and therefore we do not get any powers of $C_{q+1}^{\mu_1\cdots\mu_{q+1}}$ higher than three. Actually, due to the relation (\ref{c2p}), the theta parameters will be 
linear in $C_{q+1}^{\alpha_1\cdots\alpha_{q+1}}$
\beq
\Theta^{\alpha_{1}\cdots \alpha_{q+1}}_{\widetilde{{\sss {\rm OD}q}}}=
-(\alpha')^{\frac{q+1}{2}}(1+\frac{1}{(q+1)!}C_{q+1}^{2})^{-1}
C_{q+1}^{\alpha_1\cdots\alpha_{q+1}}\ .
\eeq

As a test we now insert the magnetically RR deformed D$(q+2)$-brane solution
 (\ref{dm}) in (\ref{ODqtheta2}), followed by taking a 
magnetic near horizon limit \cite{soloper}. This gives 
\begin{equation}
\Theta^{2\cdots (q+2)}_{\widetilde{{\sss {\rm OD}q}}}=-\theta\ell^{q-1}\ ,
\end{equation}
where $\theta$ has dimension $({\rm length})^{2}$ and $\ell$ has dimension 
length. This result for the 
`magnetic' theta, \ie the theta parameter for a magnetically deformed brane
 solution, is exactly the same as for the generalized non-commutativity 
parameter $\theta_{{\sss {\rm D}q{\rm -GT}}}$ for the non-commutative 
generalized gauge 
theories (D$q$-$GT$) defined in \cite{soloper} 
(except for a trivial sign 
difference). This was expected since the 
supergravity dual of D$q$-GT is given by the magnetic near horizon limit of 
the (D$(q+2),F1$) bound state, obtained by a magnetic RR $(q+1)$-form 
deformation of a D$(q+2)$-brane.

Next, we will check (\ref{ODqtheta2}) for supergravity solutions which 
corresponds to NS5-branes with electric or magnetic RR ($p+1$)-form 
deformations $p\neq 2$. Inserting (\ref{NS5e}) and (\ref{NS5m}) in 
(\ref{ODqtheta2}) and taking electric 
and magnetic near horizon limits (see (20) in \cite{soloper}) gives
\begin{equation}\label{the}
\Theta^{0\cdots p}_{{\sss {\rm OD}p}}=-\ell_{{\sss {\rm OD}p}}^{p+1}\ ,
\end{equation}
in the electric cases and 
\begin{equation}\label{thm}
\Theta^{(5-p)\cdots 5}_{{\sss {\rm OD}p}}=\theta\hat{\ell}^{q-1}\ , 
\end{equation}
in the magnetic cases. As expected, (\ref{thm})  agree with 
$\theta_{\rm{D}q{-\widetilde{\rm GT}}}$ in \cite{soloper}. Furthermore, just
 as in the 
string case, the theta parameter (\ref{the}) of the light D$p$-branes is 
given by the length scale. 

In the special case $p=2$, we have to use the open membrane generalized theta
 parameter (\ref{OMtheta})  below (change $\ell_p^{3}$ to $(\alpha')^{3/2}$), 
since both $C_{012}$ and $C_{345}$ are non-zero due to the duality relation. 
This gives
\begin{equation}
\Theta^{012}_{{\sss {\rm OD}2}}=-\Theta^{345}_{{\sss {\rm OD}2}}=
-\ell_{{\sss {\rm OD}2}}^{3}\ ,
\end{equation} 
and the theta parameter is again given by the length scale.

\section{The open membrane metric and generalized theta parameter}

In this section we will derive the open membrane metric and
 generalize the two-form non-commutativity
 parameter appearing in the string case to a three-form, which will be 
referred to as the open membrane theta parameter\footnote{The three form theta
 parameters discussed here and elsewhere \cite{us2,others} can not,
as in the string case, appear in
 the two point function but might arise in three point functions and in
 operator product expansions of two $x^{\mu}$'s. Therefore, strictly speaking, 
the theta parameter should probably be defined with upper indices and be 
referred to as a trivector in analogy with the bivector in the string case; 
see, \eg\  \cite{catfelderreview}.}. We will again assume deformation
independence, now of the open membrane metric.  The deformed solution we will
use is the (M5,M2) solution, from which the supergravity dual of OM-theory can be constructed by taking the appropriate near horizon limit (see (\ref{oml}) below).
Note that here, 
in contrast to the string case \cite{Berman,Delle},
there is no known method that gives rise to a guaranteed 
deformation independent open membrane metric. Instead we will use 
duality relations between solutions to argue for the correct solution to
start from.

The particular form of the (M5,M2) bound state solution 
we will make use of here is, in the five-brane directions, given by

\begin{eqnarray}\label{M5M2}
ds^{2}_{1+5}&=&H^{-\frac{1}{3}}h^{-\frac{2}{3}}\Big((dx^{0})^{2}+
(dx^{1})^{2}+(dx^{2})^{2}\Big)+H^{-\frac{1}{3}}h^{\frac{1}{3}}\Big((dx^{3})^{2}+(dx^{4})^{2}+(dx^{5})^{2}\Big)\
, \nonumber\\
A_{012}&=&\frac{\theta}{Hh}\ ,\quad
A_{345}=-\frac{\theta}{H}\ ,\quad
 h=1-\theta^{2}H^{-1}\ ,\quad H=1+\frac{N\ell_{p}^{3}}{r^{3}}\ .
\end{eqnarray}
This solution can be obtained by lifting the (D4,F1) solution in the particular
form given in
 (1) of
\cite{soloper} to eleven dimensions. That solution is written in the proper 
coordinates to yield  deformation independent 
open string metric and coupling constant, and we therefore expect that lifting
 that solution to 11 dimensions should provide a solution that can be used 
to obtain a deformation independent open membrane metric. 
For $\theta=0$ the configuration is equal to the
undeformed five-brane, and for $\theta=1$ it defines the OM dual with
critical scaling in the asymptotic region where $H\rightarrow 1$.

The open membrane metric $G^{{\sss \rm OM}}_{\mu\nu}$ is by definition a
covariant function of a slowly varying background. 
Requiring the open membrane metric to be deformation independent implies that
 it should 
be equal to the undeformed metric
\beq
g_{\mu\nu}\vert_{\rm{undeformed}}=H^{-{1\over 3}}\eta_{\mu\nu}\ .\la{nondef}
\eeq 
We make the following Ansatz for the open membrane metric 

\beq \label{OMmetricansatz}
G^{{\sss \rm OM}}_{\mu\nu}=D(K^{2})\left(g_{\mu\nu}+{1\over 4}
(A^2)_{\mu\nu}\right)\ ,\quad K=\sqrt{1+{1\over 24}A^2}\ ,\eeq
where
\begin{equation}
(A^2)_{\mu\nu}=g^{\mu_{1}\nu_{1}}g^{\mu_{2}\nu_{2}}A_{\mu_{1}\mu_{2}\mu}
A_{\nu_{1}\nu_{2}\nu}\ ,\quad A^{2}=g^{\mu\nu}(A^{2})_{\mu\nu}\ .
\end{equation}
The form of the tensorial factor can be argued for by noticing that it is this
 particular combination of $g_{\mu\nu}$ and $A_{\mu\nu\rho}$ that appears 
in the field equations on the M5-brane \cite{Howe1, Gibbonswest}. The conformal factor, 
on the other hand, is harder to derive and it has previously been obtained to
 leading order in \cite{us2}. However, more recently, 
 in \cite{janpieter} its complete form was derived by reducing the 
open membrane metric to five 
dimensions and comparing it to the known answer for the open string metric.
 One of the main 
purposes of this paper is to derive the complete form of the open membrane metric 
without resorting to reductions to five dimensions. The derivation presented 
below therefore gives further and independent evidence for the correctness of
the form of the open membrane metric obtained here and in \cite{janpieter}.

Inserting (\ref{M5M2})  into (\ref{OMmetricansatz}) and equating the OM metric 
 to the expression in (\ref{nondef}) gives the following equation:
\begin{equation}\label{eq1}
D(K^{2})H^{-1/3}h^{-2/3}\Big(\frac{h+1}{2}\Big)=H^{-1/3}\ ,
\end{equation}
where
\begin{equation}\label{eq1b}
K^{2}=h^{-1}\Big(\frac{h+1}{2}\Big)^{2}\ ,
\end{equation}
and hence the function $D(K^{2})$ is given by
\begin{equation}\label{eq2}
D(K^{2})=h^{2/3}\Big(\frac{h+1}{2}\Big)^{-1}\ .
\end{equation}
Instead of now trying to find an expression
for $D(K^{2})$, it is easier to first obtain an expression for $\hat{D}(K^{2})$
 defined as
\begin{equation}\label{eq1c}
\hat{D}(K^{2})=[D(K^{2})K]^{3}=h^{1/2}\ .
\end{equation}
Using (\ref{eq1b}), one finds the following expression for $\hat{D}(K^{2})$:

\begin{equation}\label{Bhat}
\hat{D}(K^{2})=K(1-\sqrt{1-K^{-2}})\ ,
\end{equation}
where we should note that the 
minus sign in front of the second term must be chosen, since  $h<1$ for all
 $r$. Thus, extracting $D(K^2)$ from (\ref{eq1c}) and  (\ref{Bhat}) 
 gives the following open membrane metric 

\begin{equation}\label{OMmetric}
G_{\mu\nu}^{{\sss {\rm OM}}}=\Big(\frac{1-\sqrt{1-K^{-2}}}{K^{2}}\Big)^{1/3}
\Big(g_{\mu\nu}+\frac{1}{4}(A^{2})_{\mu\nu}\Big)\ ,
\end{equation}
which may  also be written as 
\begin{equation}\label{OMmetric1}
G_{\mu\nu}^{{\sss {\rm OM}}}=[K^{4}(1+\sqrt{1-K^{-2}})]^{-1/3}\Big(g_{\mu\nu}+
\frac{1}{4}(A^{2})_{\mu\nu}\Big)\ ,
\end{equation}
or
\begin{equation}\label{OMmetric2}
G_{\mu\nu}^{{\sss {\rm OM}}}=\frac{[(2K^{2}-1)-2K^{2}\sqrt{1-K^{-2}}]^{1/6}}
{K}\Big(g_{\mu\nu}+
\frac{1}{4}(A^{2})_{\mu\nu}\Big)\ .
\end{equation}
These expressions are valid for a generic three form $A_{\mu\nu\rho}$
in contrast to the open D$q$ metrics discussed in the previous section which
are valid only for field configurations corresponding to 
one-parameter deformations.
If we compare (\ref{OMmetric2}) for the open membrane metric with the open 
membrane metric (3.28) in \cite{janpieter}, which was obtained using a 
completely different method, we see that the two expressions are identical.
We therefore have two \emph{independent} methods which give the same result,
 strongly indicating the correctness of the result, but
 the ultimate test is of course to obtain
 (\ref{OMmetric}) from a microscopic formulation of OM theory, similar to 
how the open string metric has been obtained.  

In the remainder of this section we will discuss the possibility to define
 a three form theta parameter by generalizing the 
open string two form non-commutativity parameter to the open membrane case
\footnote{For an independent discussion of theta parameters for the open membrane case, see \cite{ericjanpieter}. }.
 To be physically acceptable, we demand  
the six dimensional three-form theta parameter to be constant , \ie 
independent of $r$, when evaluated in terms of the solution  (\ref{M5M2}), 
analogously to what happens in the case of the open string theta parameter, 
where it corresponds to theta being RG flow independent
\cite{intr,peet,Berman}.
Using this assumption and (\ref{M5M2}) we find the following three-index 
theta parameter:

\begin{equation}\label{OMtheta}
\Theta^{\mu\nu\rho}_{{\sss {\rm OM}}}=-\ell_{\rm p}^{3}
[K(1-\sqrt{1-K^{-2}})]^{2/3}g^{\mu\mu_{1}}
A_{\mu_{1}\nu_{1}\rho_{1}}G^{\nu_{1}\nu}_{{\sss {\rm OM}}}
G^{\rho_{1}\rho}_{{\sss {\rm OM}}}
\ .
\end{equation}
whose form is unique in the sense that no other combinations of 
$g^{\mu\nu}$ and $G_{{\sss {\rm OM}}}^{\mu\nu}$
would generate an acceptable result under the conditions stated above.
Another way to see that there is only one acceptable form in this case is to 
note, that we can not rewrite this theta by exchanging the open membrane brane
 metric with the ordinary metric as we did for the theta parameter 
(\ref{ODqtheta2}) in the case of open D-branes. The reason for this is 
the nonlinear self-duality condition for $A_{\mu\nu\rho}$, 
yielding components in both of the blocks in the SO(1,5)/SO(1,2)$\times$SO(3) 
parametrization (see below). Furthermore, the theta parameter above is the only one reducing correctly to the open string and open D2-brane theta parameters,  as will be shown in section 5.

A consequence of the nonlinear self-duality condition satisfied by 
$A_{\mu\nu\rho}$ is, that any theta constructed from it must itself satisfy
 some self-duality  condition. In fact, the physics on the five brane should be
describable in terms of only the open membrane quantities. As such one may replace 
$g_{\mu\nu}$ and $A_{\mu\nu\rho}$ with $G^{{\sss {\rm OM}}}_{\mu\nu}$ and 
$\Theta_{{\sss {\rm 
OM}}}^{\mu\nu\rho}$. The self-duality must then be expressible using only the 
appropriate metric, $G^{{\sss {\rm OM}}}_{\mu\nu}$. 
Therefore, one might suspect that the three form theta
  satisfies a linear self-duality condition with respect to
the open membrane metric. Interestingly enough, enforcing such a linear 
condition leads exactly to the theta parameter with
 two explicit open membrane metrics as in the definition given above. 
Other definitions 
therefore can not satisfy this kind of simple
self-duality condition. One way to make this plausible, is to consider the
 electric near horizon limit of (\ref{M5M2}), \ie the $\ell_{\rm p}\rightarrow
 0$ limit 
keeping 
the following quantities fixed  
\begin{equation}\label{oml}
\tilde{x}^{\mu}=\frac{\ell_{{\sss {\rm OM}}}}{\ell_{\rm p}}x^{\mu}\ , 
\quad \tilde{r}=\frac{\ell_{{\sss {\rm OM}}}}{\ell_{\rm p}}r\ , 
\quad \theta=1\ , \ell_{{\sss {\rm OM}}}\ .
\end{equation}
Taking this limit, we obtain the following 
open membrane metric and theta parameter
\begin{eqnarray}\label{OMq}
\frac{G_{\mu\nu}^{{\sss {\rm OM}}}}{\ell_{\rm p}^{2}}&=&
\frac{1}{\ell_{{\sss {\rm OM}}}^{2}}
H^{-\frac{1}{3}}\ ,\nonumber\\
\Theta^{012}&=&\ell_{{\sss {\rm OM}}}^{3}\ ,\\
\Theta^{345}&=&\ell_{{\sss {\rm OM}}}^{3}\ ,\nonumber
\end{eqnarray} 
and hence this theta parameter is 
linearly self-dual with respect to the open membrane metric in this limit.
 Note that also in the case of open membranes, the generalized theta 
parameter is given by the intrinsic length scale of the theory.

We end this section by proving  that the theta defined in (\ref{OMtheta}) 
does indeed satisfy a linear self-duality condition with respect to 
the open membrane metric given in (\ref{OMmetric}). 
 To do this we use a frame 
$(u^{\a}_{\mu},v^a_{\mu})$ where $\a=0,1,2$ and $a=3,4,5$ 
parametrized by the coset 
SO(1,5)/SO(1,2)$\times$SO(3) 
as defined in \cite{us2}. With this parametrization we can write a generic
 three 
form $A_{\mu\nu\rho}$  in terms of one function as
\beq\label{Athreethreesplit}
A_{\mu\nu\rho}=a_1u^3_{\mu\nu\rho}+a_2v^3_{\mu\nu\rho}\ ,
\eeq
where, due to the non-linear self-duality \cite{us2},
$a_1=a_2/\sqrt{1+a_2^2}$.   
Then a general open membrane metric can be expressed as
\begin{eqnarray}\label{ommetric}
G^{{\sss {\rm OM}}}_{\alpha\beta}&=&\Big(1+\frac{1}{6}A_{1}^{2}\Big)^{2/3}
g_{\alpha\beta}\ ,
\quad A_{1}^{2}=g^{\alpha\beta}(A_{1}^{2})_{\alpha\beta}\ ,\nonumber\\ 
(A_{1}^{2})_{\alpha\beta}&=&g^{\alpha_{1}\beta_{1}}g^{\alpha_{2}\beta_{2}}
A_{\alpha_{1}\alpha_{2}\alpha}
A_{\beta_{1}\beta_{2}\beta}\ ,\quad \alpha,\beta=0,1,2\ ,\nonumber\\
G^{{\sss {\rm OM}}}_{ab}&=&\Big(1+\frac{1}{6}A_{2}^{2}\Big)^{1/3}g_{ab}\ ,
\quad A_{2}^{2}=g^{ab}(A_{2}^{2})_{ab}\ ,\\
(A_{2}^{2})_{ab}&=&g^{a_{1}b_{1}}g^{a_{2}b_{2}}
A_{a_{1}a_{2}a}
A_{b_{1}b_{2}b}\ ,\quad a,b=3,4,5\ ,\nonumber
\end{eqnarray}
where we have used that
\begin{eqnarray}
K^{2}&=&\frac{(1+\frac{1}{12}A_{i}^{2})^{2}}{1+\frac{1}{6}A_{i}^{2}}\ ,\quad 
i=1,2\ ,\nonumber\\
1+\frac{1}{6}A^{2}_{2}&=&\Big(1+\frac{1}{6}A^{2}_{1}\Big)^{-1}\ .
\end{eqnarray}
The last two relations follow directly from the relation between 
$a_1$ and $a_2$
given above, remembering that $u^3_{\mu\nu\rho}$ squares to $-$6 while 
$v^3_{\mu\nu\rho}$ squares to +6 (for more details see \cite{us2}).
The theta parameter can then be written as
\beq\label{thetaabc}
\Theta^{\alpha\beta\gamma}=(1+\frac{1}{6}A_2^2)\,A^{\alpha\beta\gamma}\,,\qquad
\Theta^{abc}=(1+\frac{1}{6}A_1^2)\,A^{abc}\ ,
\eeq
and be shown to satisfy
\beq
\ast_G \Theta = \Theta\ ,\label{duality}
\eeq
where the Hodge dual is defined as\footnote{The Levi-Civita tensor is 
here defined with $\epsilon^{01\cdots 5}=1$.}
\beq \label{sd}
(\ast_G \Theta)^{\mu\nu\rho}=\frac{1}{6}\frac{1}
{\sqrt{-G}}\epsilon^{\mu\nu\rho\iota\kappa\lambda}\Theta_{\iota\kappa\lambda}\ .
\eeq
Here $G$ is the determinant of the open membrane metric and the 
indices on $\Theta$ are 
lowered with $G^{{\sss {\rm OM}}}_{\mu\nu}$. We therefore see that 
$\Theta^{\mu\nu\rho}$ is linearly self-dual with respect to the 
open membrane metric $G^{{\sss {\rm OM}}}_{\mu\nu}$. We also see from (\ref{thetaabc}) 
that $\Theta^{\mu\nu\rho}$ is completely anti-symmetric.

\section{Dimensional reduction of open membrane data}

In this section we show that we can obtain the data for the open 
D2-brane corresponding to the $\widetilde{\mathrm{OD}2}$-theory 
 as well as the open string data corresponding to NCOS by dimensional
 reduction of the open membrane data, \ie\ the open membrane metric and 
theta parameter. When making these relations precise, we will find that only
 under certain conditions can the five-dimensional data be lifted to six 
dimensions. In particular, the open string metric with $B^2\geq 0$, 
which we are accustomed to use
 in a magnetic limit to obtain NCYM, does not seem to have any direct connection to the six
 dimensional open membrane metric discussed in the previous section. The 
electric case with $B^2\leq 0$ related to NCOS does, on the other hand, lift as expected.

We start from the open membrane data  expressed in terms of a background 
metric and a three-form, which obeys a non-linear 
self-duality constraint \cite{Howe1}. 
When reducing from six to five dimensions, we obtain  from the three-form 
$A_{\hat{\mu}\hat{\nu}\hat{\rho}}$
 (hatted indices are six dimensional in this section) both a
two-form and a three-form which, however, 
are related due to the self-duality constraint.  
Here we will work with the supergravity fields, which according to the standard
conventions reduce as follows (the far left hand sides of these equations are 
components of six dimensional quantities: $\hat{\mu}=(\mu,y)$):
\bea
\frac{g_{\mu\nu}}{\ell_{\mathrm{p}}^2}&=&e^{-\frac{2\phi}{3}}\,
\frac{g^s_{\mu\nu}}{\alpha'}=\frac{g^{{\sss \mathrm{D}2}}_{\mu\nu}}{\alpha'}
\,,\qquad\frac{g_{yy}}{\ell_{\mathrm{p}}^2}=\frac{e^{\frac{4\phi}{3}}}{R^2}
\label{gg}\,,\\
\frac{A_{\mu\nu\rho}}{\ell_{\mathrm{p}}^3}&=&\frac{C_{\mu\nu\rho}}
{\alpha'^{\frac{3}{2}}}\,,\qquad\frac{A_{\mu\nu y}}{\ell_{\mathrm{p}}^3}=
\frac{B_{\mu\nu}}{\alpha'R}\,,\label{ABC}
\eea
where $R$ is the radius of the compactified direction labeled by $y$. 
The procedure is exactly the same as if we had chosen to reduce world 
volume fields; for instance 
the relations between the two-form and the three-form found in \cite{janpieter} 
hold, with a few minor modifications such
 as the inclusion of dilaton factors, for the supergravity fields as well. 

The dimensional reduction is performed for the open membrane 
\beq
G^{{\sss \mathrm{OM}}}_{\hat{\mu}\hat{\nu}}=z^{-1}K^{-1}\Big(g_{\hat{\mu}
\hat{\nu}}+\frac{1}{4}A^2_{\hat{\mu}\hat{\nu}}\Big)\,,
\eeq
where
\beq
z^{-1}=\big(K(1-\sqrt{1-K^{-2}})\big)^{\frac{1}{3}}\,,
\eeq
and also for the theta parameter 
\begin{equation}
\Theta^{\hat{\mu}\hat{\nu}\hat{\rho}}_{{\sss \rm{OM}}}=
-\ell_{\mathrm{p}}^3\,z^{-2}g^{\hat{\mu}\hat{\mu}_1}
A_{\hat{\mu}_1\hat{\nu}_1\hat{\rho}_1}\,G_{{\sss \mathrm{OM}}}^{\hat{\nu}
\hat{\nu}_1}\,G_{{\sss \mathrm{OM}}}^{\hat{\rho}\hat{\rho}_1}\ ,
\end{equation}
both discussed in detail in the previous section.
Using the above reduction formulae, and the standard parameter relations 
$\ell_{\rm{p}}^2=g^{\frac{2}{3}}\alpha'$ and $R=g\sqrt{\alpha'}$, we get
\bea
A^2_{\mu\nu}&=&g^{\frac{2}{3}}\Big(C^2_{\mu\nu}+2\,e^{-\frac{2\phi}{3}}
B^2_{\mu\nu}\Big)\,,\label{A2mn}\\
A^2&=&C^2+3B^2\,,\label{A2}
\eea
where contractions of $A, B$ and $C$ are done with $g_{\mu\nu}$, 
$g^s_{\mu\nu}$ and $g^{{\sss \mathrm{D}2}}_{\mu\nu}$, respectively. The 
choices of these particular metrics in the contractions of $B_{\mu\nu}$ and 
$C_{\mu\nu\rho}$ are dictated by the fact, that they will be used only in 
the context where they are relevant, \ie\ the former in the 
open string context and the latter in the open D2-brane context.
{}From the non-linear self-duality relation in six dimensions we get as in
 \cite{janpieter},
provided we restrict ourselves to rank two fields in five dimensions, the
 following relations between the two-form and the three-form: 
\bea
e^{-\frac{2\phi}{3}}B^2_{\mu\nu}&=&{\scs \frac{1}{6}}(1+{\scs \frac{1}{6}}\,
C^2)^{-1}\Big(3\,C^2_{\mu\nu}-C^2\,g^{{\sss \mathrm{D}2}}_{\mu\nu}\Big)
\,,\label{B2mn}\\
1+{\scs \frac{1}{2}}\,B^2&=&(1+{\scs \frac{1}{6}}\,C^2)^{-1}\,.\label{B2}
\eea

We will start by reducing on an electric circle defined by 
having a non-zero $B_{01}$ and hence $B^2<0$, 
which for a generic supergravity solution 
corresponds to a reduction from (M5,M2) to (D4,F1) under the rank two 
restriction
(rank four is a two parameter deformation, corresponding to a more complicated 
bound state, see e.g. \cite{Delle}). 
In the UV limit of 
these solutions, when interpreted as supergravity duals, 
we get light membranes in the (M5,M2) case, \ie\ OM-theory, and light strings
in the (D4,F1) case, corresponding to NCOS. This means that in this case, 
the open membrane metric should reduce to the open string metric, and we 
therefore express the 
reduced metric in terms of the two-form $B_{\mu\nu}$ with  $B^2<0$. 
In writing out the conformal 
factor in terms of the two-form, it turns out that the sign of $B^2$ 
(and of $C^2$ when using the three-form, since $B^2$ and $C^2$ always have opposite
 signs) becomes important. To be specific, the 
form of $z$ differs, when expressed in terms of the five-dimensional fields, 
for different signs of $B^2$, whereas the form of $K$ is the same for both 
signs. This difference is due to the second order equation that follows from
 (\ref{A2mn}) and (\ref{B2mn}), relating
 $B^2$ (or $C^2$) and $A^2$. For the electric reduction we get
\bea
z&=&(1+{\scs \frac{1}{2}}\,B^2)^{-\frac{1}{6}}=
(1+{\scs \frac{1}{6}}\,C^2)^{\frac{1}{6}}\,,\label{z1}\\
K&=&(1+{\scs \frac{1}{4}}\,B^2)(1+{\scs \frac{1}{2}}\,B^2)^{-\frac{1}{2}}=
(1+{\scs \frac{1}{12}}\,C^2)(1+{\scs \frac{1}{6}}\,C^2)^{-\frac{1}{2}}\,.
\label{K}
\eea
Using (\ref{gg})-(\ref{K}), the open membrane metric reduces to
\beq
\frac{G^{{\sss \mathrm{OM}}}_{\mu\nu}}{\ell_{\mathrm{p}}^2}= 
\frac{g^{\frac{2}{3}}}{\ell_{\mathrm{p}}^2}e^{-\frac{2\phi}{3}}
(1+{\scs \frac{1}{2}}\,B^2)^{-\frac{1}{3}}\Big(g^s_{\mu\nu}+B^2_{\mu\nu}\Big)\,.
\label{Gom}
\eeq
In analogy 
with (\ref{gg}) for the metric along the compactified direction, we can define the effective open string coupling from the 
open membrane metric
\beq
\frac{G_{yy}^{{\sss {\rm OM}}}}{\ell_{{\rm p}}^2}=
\frac{(G_{{\sss \mathrm{OS}}}^2)^{\frac{4}
{3}}}{R^2}\,,\label{Gy}
\eeq
which yields
\beq
G_{{\sss \rm{OS}}}^2=e^\phi(1+{\scs \frac{1}{2}}\,B^2)^{\frac{1}{2}}\,,
\label{Gos}
\eeq
which is the expected coupling in the open string case
 (analogous to the one found in 
\cite{janpieter}). We then define the open string metric as
\beq
\frac{G^{{\sss \rm{OM}}}_{\mu\nu}}{\ell_{\rm{p}}^2}=(G_{{\sss \rm{OS}}}^2)^{-\frac{2}
{3}}\frac{G^{{\sss \rm{OS}}}_{\mu\nu}}{\alpha'}\,,\label{Gosmn}
\eeq
also in analogy to the reduction formulae for the closed string quantities
 (\ref{gg}).
Inserting (\ref{Gom}) and (\ref{Gos}) into (\ref{Gosmn}) we get the expected 
open string metric
\beq
G^{{\sss \rm{OS}}}_{\mu\nu}=g^s_{\mu\nu}+B^2_{\mu\nu}\,.
\eeq
If one wants to relate open membrane and open string quantities without
 involving 
any closed quantities, one is naturally led to the above definitions. Thus by
requiring deformation independence in six dimensions we can uniquely determine
 the
open membrane metric and its connection via dimensional reduction to the open
 string metric.
This argument can be turned around as done in \cite{janpieter}, where the
 fundamental
assumption instead concerns the relations used in the dimensional reduction 
of the open membrane metric (\ref{Gy}) and (\ref{Gosmn}).

The expression for the open string metric was derived 
by performing an electric reduction, \ie having $B^2<0$, but 
it can be analytically continued to $B^2>0$, which is needed in order 
to get  NCYM in the decoupling limit. 
 We have thus obtained the open string metric by 
electrically reducing the open membrane metric and then 
analytically continuing the obtained expression to all values of $B^2$. 
However, although it is generally considered a standard procedure, 
there seems to be an obstacle  to lifting the five dimensional
open string metric to six dimensions for all values of $B^2$. The problem 
stems from the relation between the five dimensional quantity 
$B^2$ and and the six dimensional one $K=\sqrt{1+\frac{1}{24}A^2}$
 discussed in the previous section. Using the relations between the 
fields in five and six dimensions, one finds the relation 
$B^2=4(K^2-1\pm K^2\sqrt{1-K^{-2}})$ where $K>1$. Thus $B^2>0$ corresponds 
uniquely
to the plus sign and $B^2<0$ to the minus sign. But in
the previous section (see comment below (\ref{Bhat})) we showed that the minus 
sign is the only possible one, in agreement with 
the result of \cite{janpieter}. 
Our interpretation of this is that in the magnetic case with $B^2>0$ one
 should not use the open string metric to analyze the physics but rather the 
open D2-brane metric given in (\ref{ODqmetric}). Thus, as we will see below, 
it is in this case the open
D2-metric that  lifts naturally to six dimensions.

Having done the reduction on an electric circle, we now turn to the reduction
 on a magnetic circle, which for a generic solution means that (M5,M2) 
reduces to (D4,D2) under the restriction to rank two in five dimensions. 
We now have an electric three-form, \ie $C^2<0$, and therefore also a magnetic
 two-form, \ie $B^2>0$. The difference compared to the electric case above is
 that now
\beq
z=(1+{\scs \frac{1}{2}}\,B^2)^{\frac{1}{6}}=(1+{\scs \frac{1}{6}}\, 
C^2)^{-\frac{1}{6}}\label{z2}\,.
\eeq
The procedure is exactly as above, but since light membranes now 
reduce to light D2-branes, we expect to obtain the open D2-brane metric
 (\ref{OD23}), and we should therefore express the reduced metric in terms 
of the three-form. Indeed, for the magnetic reduction we find
 the result 
\bea
\frac{G^{{\sss \mathrm{OM}}}_{\mu\nu}}{\ell_{\mathrm{p}}^2}=
\frac{1}{\alpha'}(1+{\scs \frac{1}{6}}\,C^2)^{-\frac{1}{3}}
\Big(g^{{\sss \rm{D}2}}_{\mu\nu}+{\scs \frac{1}{2}}\,C^2_{\mu\nu}\Big)=
\frac{G^{{\sss \widetilde{\rm{OD}2}}}_{\mu\nu}}{\alpha'}\,.\label{God}
\eea
As in the previous case we have to analytically continue the 
open D2-brane metric to the case $C^2>0$ relevant for D2-GT. 
One could speculate that $G^{{\sss \mathrm{OM}}}_{yy}$ in the above case is
 related to the open D2-brane coupling. However, as we have seen above, 
(\ref{God}) can be obtained without knowing the exact relation between 
$G^{{\sss \mathrm{OM}}}_{yy}$ and the open D2-brane coupling.

We now turn to the reduction of the eleven dimensional theta parameter. 
Using the reduction rules and the
 definitions above, we get after an electric reduction
\beq\label{OStheta1}
\Theta^{\mu\nu}_{\sss \rm{OS}}\equiv\frac{\Theta^{\mu\nu y}_{{\sss 
\rm{OM}}}}{R}=-\alpha'\,
g_s^{\mu\mu_{1}}B_{\mu_{1}\nu_{1}}G_{{\sss \rm{OS}}}^{\nu\nu_{1}}\ ,
\eeq
and the theta parameter of OM-theory therefore reduces to the correct open string non-commutativity parameter. Instead, reducing on a magnetic circle we 
obtain the open D2-brane three-index theta
\beq\label{ODtheta1}
\Theta^{\mu\nu\rho}_{\sss\widetilde{\rm{OD}2}}=
\Theta^{\mu\nu\rho}_{\sss\rm{OM}}=-(\alpha')^\frac{3}{2}(1+{\scs \frac{1}{6}}\, 
C^2)^{\frac{1}{3}}g^{\mu\mu_1}_{\sss\rm{D}2}C_{\mu_1\nu_1\rho_1}
G^{\nu\nu_1}_{\sss\widetilde{\rm{OD}2}}G^{\rho\rho_1}_{\sss\widetilde{\rm{OD}2}}\,.
\eeq
Importantly the open string and open D2-brane theta parameters can be obtained in both the electric and magnetic reductions of (\ref{OMtheta}), whereas an open membrane theta parameter defined in another way will reduce to two 2-index and two 3-index parameters. That we only get one 2-index and one 3-index theta, can be 
understood from the duality relation (\ref{duality}). Since all $z$ 
dependence in the eleven-dimensional Hodge star cancels, 
it reduces in the same way in both the electric and magnetic case. 
Since it relates physical quantities in eleven dimensions, this should also 
be the case in ten dimensions, implying that we can only get {\em one} 
expression for the two-index theta and {\em one} for the three-index theta in ten
 dimensions. We have thus seen that the open membrane data correctly reduce to 
 the open string and D2-brane data. In this way 
we can also relate the various decoupling limits  including the NCYM and D2-GT 
cases.
Note also that trying to obtain the open D2-brane metric from an electric 
reduction or an open string metric from the magnetic reduction will fail. 

Finally, we note that if we reduce the open membrane data in a direction transverse to the M5-brane (\ie (M5,M2) goes to (NS5,D2)), we get exactly the same expressions for the open 2-brane (with a self-dual three-form) metric and theta parameter, but with the identifications between the metrics and the three-forms (\ref{gg}).

\section{Summary and discussion}
In this paper we derive expressions for open brane metrics and theta parameters
for open D$q$-branes ending on NS5-branes or D$(q+2)$-branes, and for
the open membrane ending on an M5-brane. For the open string these quantities
can be read off from the two-point function, which is easily computed
by quantizing the open string in the background of a constant metric and 
NS-NS field $B_{\mu\nu}$. The open string quantities so obtained
constitute the data needed in taking certain closed string decoupling limits
and thus contain the information required to define new interesting theories
like NCYM and NCOS. 

In order to generalize these 
open string quantities to the open branes mentioned above, 
 we have to rely on some indirect methods since
in these cases there is no  known method to compute 
two- or higher-point functions 
from basic principles. To proceed, one could imagine that the open brane quantities could be derived through S- and T-duality transformations, where the former may involve a dimensional lift, \eg\ from five to six dimensions as
discussed in this paper. Now, applying these transformations
leads to ambiguities as to how the open scalar quantities (like the conformal factor
and open brane coupling) after the transformations
should be defined. This ambiguity can be eliminated only by invoking a
further assumption tied to the relations between open brane quantities in 
different dimensions, or by generalizing some key properties exhibited by
the open string data and regarding them as general principles.

In the recent work \cite{janpieter} the open membrane metric was 
determined, including the conformal factor, by invoking an explicit
assumption concerning how the open string metric in five dimensions is related
to the open membrane metric in six dimensions, namely   
eqs. (\ref{Gy}) and (\ref{Gosmn}). In fact, these relations follow also by assuming
that they should only involve open string quantities.

In this paper we have instead chosen to elevate the facts that the open string metric
is deformation independent and that $\Theta^{\mu\nu}$ is scale independent to general
principles. This has the advantage that the argument can be applied to each case separately
and may therefore have a more direct physical explanation. It also simplifies the 
computations considerably making the results more transparent. We have shown that the two approaches, relying on deformation independence and reduction respectively, yield the same open membrane metric. 

The ultimate reason for the interest in these generalized open brane quantities
is of course related to the possibility, that they contain information
about the quantum theory in cases where we do not know how to extract such 
information by other means. In particular, the appearance of generalized 
theta parameters may indicate new algebraic structures, arising in operator
 products of fields generated by the open branes
on the host branes. Related speculations have already appeared in the literature,
 \eg\ in \cite{others}, and to find any kind of explicit realization of such 
structures is clearly an 
important problem.

\vspace{0.4cm}
\Large \textbf{Acknowledgments}
\vspace{0.2cm}
\normalsize

We are grateful to Jan Pieter van der Schaar and Eric Bergshoeff for 
discussions and for sharing their results with us prior to publication. 
This work is partly supported by EU contract HPRN-CT-2000-00122
and by the Swedish Research Council. DSB is funded by the BSF-American-Israeli 
bi-national science foundation, the center of excellence project, ISF and the 
German-Israeli bi-national science foundation.

\end{document}